# High order nonlinear electrophoresis in a nematic liquid crystal

Mojtaba Rajabi [1,2], Taras Turiv [1], Bing-Xiang Li [1,3], Hend Baza [1,2], Dmitry Golovaty [4], Oleg D. Lavrentovich [1,2,5,*]

[1] Advanced Materials and Liquid Crystal Institute, Kent State University, Kent, OH 44242, USA
[2] Department of Physics, Kent State University, Kent, OH 44242, USA
[3] College of Electronic and Optical Engineering & College of Flexible Electronics (Future Technology), Nanjing University of Posts and Telecommunications, Nanjing 210023, China
[4] Department of Mathematics, The University of Akron, Akron, OH 44325-4002, USA.
[5] Materials Science Graduate Program, Kent State University, Kent, OH 44242, USA
[*] Corresponding author. Email: olavrent@kent.edu

Electrophoresis is the motion of particles relative to a surrounding fluid driven by a uniform electric field. In conventional electrophoresis, the electrophoretic velocity grows linearly with the applied field. Nonlinear effects with a quadratic speed vs field dependence are gaining research interest since an alternating current field could drive them. Here we report on the giant nonlinearity of electrophoresis in a nematic liquid crystal in which the speed grows with the fourth and sixth powers of the electric field. The mechanism is attributed to the shear thinning of the nematic environment induced by the moving colloid. The observed giant nonlinear effect dramatically enhances the efficiency of electrophoretic transport.

Electrophoresis of colloidal particles has been observed in both isotropic [1-5] and anisotropic electrolytes [6-11], as reviewed recently [12,13]. In isotropic fluids, it is usually caused by the separation of charges in electric double layers around a particle [1,5,14]. An applied electric field imposes a realigning torque on the electric double layer that propels the particle. At moderate fields **E**, the electrophoretic velocity grows linearly with the field, $\mathbf{v} = \mu^{(1)}\mathbf{E}$, where $\mu^{(1)}$ is the electrophoretic mobility. At high fields, the field dependency of the electrophoretic velocity acquires a third power term, $v = \mu^{(1)}E + \mu^{(3)}E^3$ [10,15], explained by the field-induced modification of the double layer [16,17]. Although the odd-power electrophoresis is important in practical applications [18], there is a growing interest in its quadratic counterpart, $v \propto E^2$, which can be driven by an alternating current (ac) electric field. The ac driving mitigates undesirable voltage screening and chemical reactions at electrodes, allows one to create steady flows and design vortices for microscale mixing [19]. In electrophoresis with $v \propto E^2$ the spatial separation of charges is caused by the electric field. In isotropic electrolytes, the effect is called the induced-charge electrophoresis (ICEP) [19]. To be ICEP active, the particle must be asymmetric in shape or properties [20,21]. This requirement of particle asymmetry is lifted when the electrolyte represents a nematic liquid crystal (LC), a fluid with a long-range orientational order of molecules along some direction $\hat{\mathbf{n}} \equiv -\hat{\mathbf{n}}$ called the director. When a colloid is placed in a nematic, it distorts the director [22]. Because of the anisotropy of dielectric permittivity and conductivity of the LC, an applied electric field induces spatial charges at these distortions. If the director distortions around the particle lack a fore-aft symmetry, the colloid will be electrophoretically active even if it is an ideal sphere [8,23]. The effect is called liquid crystal-enabled electrophoresis (LCEP) [9]. The LCEP has been so far explored for relatively weak electric fields, when the LC electrolyte is not disturbed much by the applied field and colloid's motion [8,9,11,23].

Here we report on giant nonlinear LCEP, in which the electrophoretic speed of colloidal spheres grows as the fourth and sixth powers of the applied field. The phenomenon cannot be explained by field-induced director rearrangements since it persists in a nematic with a vanishingly small dielectric anisotropy. We attribute the effect to the shear thinning of the LC electrolyte. Motion of the colloid creates a shear in the nematic which reduces its viscosity. Independent rheological measurements confirm that the nematic viscosity decreases as the shear rate increases. The observed giant nonlinear effect is a consequence of the strongly non-Newtonian nature of the nematic electrolyte, which dramatically enhances the efficiency of electrophoretic transport; it can be useful in applications such as the separation of colloidal particles of identical shapes and bulk features but different surface properties [24].

*Materials and methods.* We use the nematic MLC7026-000 (purchased from Merck) with a negative dielectric anisotropy $\Delta\varepsilon = \varepsilon_\parallel - \varepsilon_\perp = -3.7$, and mixtures of MLC7026-000 and E7 (Merck) of weight proportions of 89.1:10.9 with a vanishing $|\Delta\varepsilon| < 10^{-3}$ and 95:5 with $\Delta\varepsilon \approx -1$. Here, $\varepsilon_\parallel$ and $\varepsilon_\perp$ are dielectric permittivities measured parallel and perpendicular to $\hat{\mathbf{n}}$, respectively. MLC7026 was previously used to describe the quadratic LCEP effect in a weak field, $E \leq 0.04\ \text{V}/\mu\text{m}$ [9]. In this work, we extend the study to higher field amplitudes, $\sim 0.8\ \text{V}/\mu\text{m}$. The clearing temperatures of the nematic mixtures are much higher than the room temperature, 80°C for MLC7026 [9] and $(76-78)$°C for the mixtures with E7. This helps to avoid a detrimental Joule effect, which is known to increase the temperature of cells similar to ours by a few degrees when the field is on the order of $1\text{V}/\mu\text{m}$ [25]. For additional verification, we performed experiments on cells in a hot stage with a cooling element which kept the temperature constant within 0.1°C; these experiments confirm the same nonlinear effect.

Cells of thicknesses 100 μm are assembled from two flat glass plates with transparent indium tin oxide (ITO) electrodes, Fig. 1(a). The plates are spin-coated with a thin layer of polyimide PI-2555 (HD MicroSystems) and rubbed unidirectionally to set planar alignment along the $x$-axis, $\hat{\mathbf{n}}_0 = (1,0,0)$. Polystyrene (PS) and glass spheres (both from Duke Scientific Corporation) of the diameter $2R = 5\ \mu\text{m}$ are treated with silane N, N-dimethyl-N-octadecyl-3-aminopropyl-trimethylsilyl chloride (DMOAP) (Sigma Aldrich) to impose homeotropic (perpendicular) surface anchoring of $\hat{\mathbf{n}}$ [26]. The homeotropic alignment causes dipolar director distortions $\hat{\mathbf{n}}(x,y,z)$ with a point defect, called a hyperbolic hedgehog, located on one side of the particle. The hedgehog connects the radial director field at the colloid's surface with the unidirectional far-field $\hat{\mathbf{n}}_0$, Fig. 1(a-c) [22]. The director field lacks the fore-aft symmetry and enables a quadratic LCEP of the colloidal spheres [8,9].

A Siglent SDG1032X waveform generator and an amplifier (Krohn-Hite corporation) are used to apply the electric field. The PS spheres do not show any significant sedimentation during the experiments since their density $\rho_{PS} = 1.05 \times 10^3\ \text{kg/m}^3$ is close to that of the LC, $\rho_{LC} \approx 10^3\ \text{kg/m}^3$. Elastic repulsion from the substrates mediated by director distortions around the colloids also helps to avoid sedimentation [27]. Whenever needed, the sedimented PS colloids are elevated into the bulk by applying a small (~10 mV/μm) direct current (dc) electric field [9]; this dc field is used only during the cell preparation and is switched off before the LCEP experiments. To counterbalance the sedimentation of the glass spheres ($\rho_G = 2.5 \times 10^3$ kg/m$^3$), we applied a very small 3 mV/μm dc electric field along the normal to the cell during the LCEP experiment. This field is 20 times weaker than the lowest driving ac field, Fig.1(d) and does not cause LCEP propulsion of $2R = 5$ μm spheres [9]. Its presence did not change the qualitative nonlinear character of LCEP.

LCEP is driven by a sinusoidal ac electric field $\mathbf{E} = (0,0,E_z)$ of the frequency $f = 20$ Hz and observed under a Nikon TE2000 optical microscope equipped with a camera CMOS (Emergent HS-20000C); the trajectories are tracked by ImageJ software. The speed of the colloids was established by determining the instantaneous positions of the center of mass of the image every 0.1 s. The experiments were performed for 3 to 6 spheres at a given fixed set of cells and material parameters.

The viscosity $\eta$ as a function of the shear rate $\dot{\gamma}$ of MLC7026 is measured using a rheometer Anton Paar, MCR-302 with a cone-plate geometry (60 mm in diameter cone and 1-degree cone angle) at room temperature (21 °C). Each measured $\eta$ point represents the steady-state response of the rheometer at the given applied shear rate. At each shear rate, the steady state is determined when the viscosities of six consecutive measurements vary by less than 2%, with each measurement averaging the response over 10 seconds. The viscosity could not be measured at rates below $\dot{\gamma} = 0.1$ s$^{-1}$. Prior experiments suggest that E7 [28] and MLC7026 [29] are flow-aligning, similarly to many low-molecular weight nematics that do not show a smectic phase [30]; the studied mixture MLC7026-000:E7 (Merck) in weight proportions of 89.1:10.9 is also flow-aligning, as follows from the comparison of the numerical simulations [31] and experiments [32] on the electro-osmotic flows. In flow aligning calamitic nematics, the Leslie coefficients $\alpha_2$ and $\alpha_3$ are both negative [30]. In particular, numerical simulations [31] with $\alpha_2 = -173$ mPa · s and $\alpha_3 = -30$ mPa · s describe well the flow behavior of the 89.1:10.9 mixture.

***Results and discussion.*** The PS spheres move *perpendicularly* to the field $\mathbf{E} = (0,0,E_z)$ along the overall director $\hat{\mathbf{n}}_0 = (1,0,0)$, confirming the LCEP mechanism [8,9]: the electric field creates space charge at the director distortions $\hat{\mathbf{n}}(x,y,z)$ around the sphere and drives the nematic flow that lacks fore-aft symmetry because of the point defect-hedgehog, Fig. 1(b,c). At a fixed $E_z = const$, the electrophoretic speed $v_x$ does not change with time, which indicates the absence of detrimental factors such as sedimentation, Supplementary Fig. S1. The speed is a strongly nonlinear function of the field, fitted with a sixth-order polynomial for $E_z < 0.4$ V/μm, Fig. 1(d):

$$v_x = \beta^{(2)}E_z^2 + \beta^{(4)}E_z^4 + \beta^{(6)}E_z^6, \quad (1)$$

where $\beta^{(2)} = 9$ (μm)$^3$/V$^2$s, $\beta^{(4)} = 150$ (μm)$^5$/V$^4$s, and $\beta^{(6)} = 22000$ (μm)$^7$/V$^6$s. At $E_z > 0.4$ V/μm the speed dependence on the field is again quadratic, Fig. 1(d). The glass spheres engage in a similar giant nonlinear electrophoresis, Fig. 1(d). Note that fields above 0.14-0.25 $V$/μm cause electrohydrodynamic instability in mixtures with $|\Delta\varepsilon| < 10^{-3}$ and $\Delta\varepsilon \approx -1$, thus the LCEP could not be characterized in these cases.

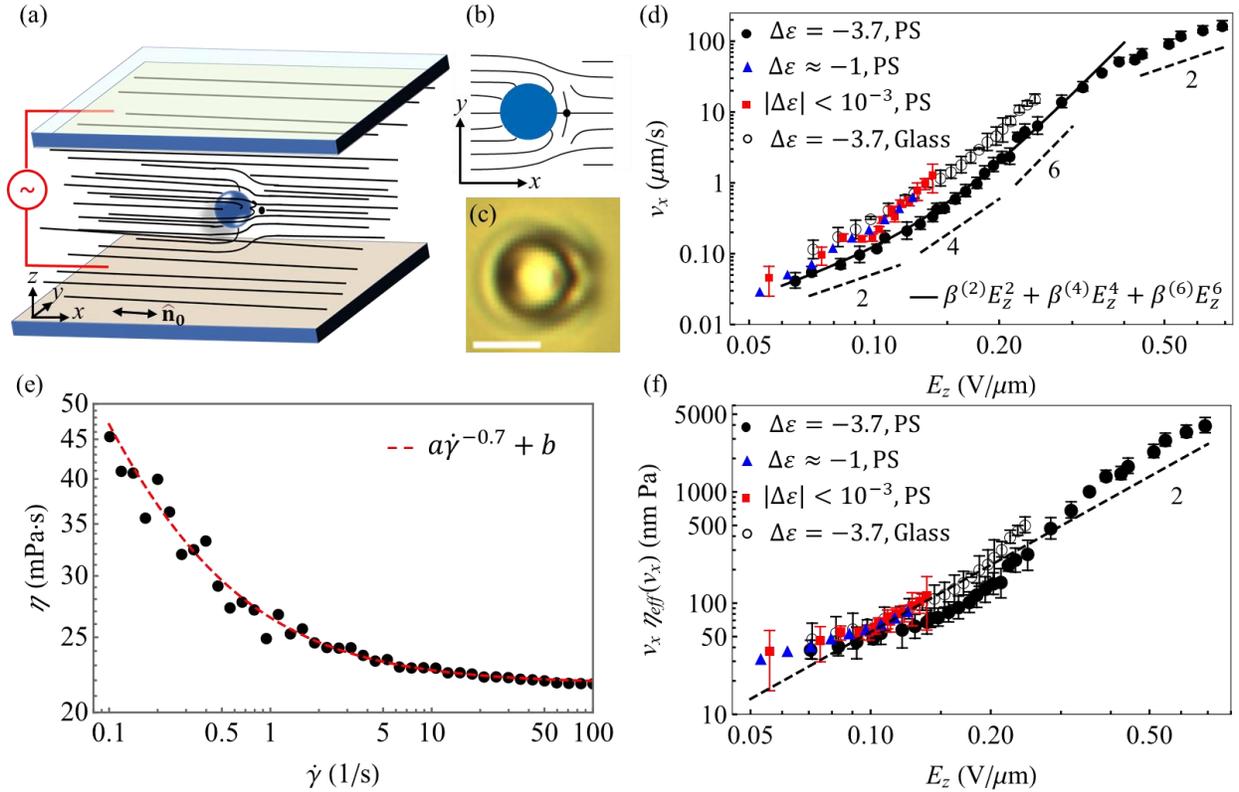

FIG. 1. Highly nonlinear electrophoresis. (a) Geometry of the experiment (not to scale). (b) Director distortions around a sphere with homeotropic anchoring. (c) Optical microscopy texture of a PS sphere with a hyperbolic hedgehog; $E_z = 0$. The scale bar is 5 μm. (d) Field dependencies of the electrophoretic speed $v_x$ of PS spheres in MLC7026, $\Delta\varepsilon = -3.7$ (closed circles), MLC7026+E7 mixtures with $\Delta\varepsilon \approx -1$ (triangles), $|\Delta\varepsilon| < 10^{-3}$ (squares), and glass spheres in MLC7026 (open circles); $f = 20\ Hz$. Error bars represent standard deviations calculated over 3-6 experiments. For the mixture with $\Delta\varepsilon \approx -1$, only one experiment has been performed. The dashed lines illustrate the slope. The solid line is a sixth-order fitting for a PS sphere in MLC7026. (e) Rheometer-measured viscosity $\eta$ of MLC7026 as a function of shear rate $\dot{\gamma}$ at 21°C. (f) Product of the speed and effective viscosity $v_x\,\eta_{eff}\,(v_x)$ as a function of electric field for LCs with different $\Delta\varepsilon$.

The LCEP speed of a colloidal sphere is expected to grow with the square of the field, being dependent on the anisotropy of conductivity $\Delta\sigma$ and permittivity $\Delta\varepsilon$, and the effective viscosity of the nematic $\eta_{eff}$ [23]:

$$v_x = b\frac{\varepsilon_0 \bar{\varepsilon}}{\eta_{eff}}\left(\frac{\Delta\sigma}{\bar{\sigma}} - \frac{\Delta\varepsilon}{\bar{\varepsilon}}\right) R E_z^2; \tag{2}$$

here $\bar{\sigma}$ and $\bar{\varepsilon}$ are the average conductivity and permittivity, respectively, and $b$ is a numerical constant that depends on the director details. Previously [8,9,11,23,33,34], all parameters in Eq. (2) have been treated as field-independent, which was a reasonable assumption for weak fields and small velocities. In strong fields this assumption is not necessarily valid.

The first obvious effect of a strong field is the change of the director configuration around the sphere, clearly seen in Fig. 2(a) for MLC7026, in which the dielectric anisotropy is high, $\Delta\varepsilon = -3.7$. However, when electrophoresis is explored in a mixture of MLC7026 and E7 with a vanishingly small $|\Delta\varepsilon| < 10^{-3}$, the observed in-plane field-induced director distortions are much weaker, Fig. 2(b), yet the giant nonlinearity is still in place, Fig. 1(d). Another argument against the importance of dielectric torques is that the effective "passive" viscosities of MLC7026, $\eta_{eff,\parallel}$ and $\eta_{eff,\perp}$ (for motion parallel and perpendicular to $\hat{\mathbf{n}}_0$, respectively) characterizing Brownian diffusion of colloids do not change in the presence of an electric field of high frequency of 1 KHz, Fig. 2(c) and Supplementary Fig. S2. At this frequency, the field distorts the director but does not cause electrophoresis of PS spheres. We conclude that the electrically-triggered "breathing" of the director around the colloid is not the main reason for the giant electrophoretic nonlinearity. A stronger effect is produced by the shear thinning of the nematic electrolyte.

A colloidal sphere moving under an applied electric field creates a shear of flow in its environment. This shear, acting on the deformed director field $\hat{\mathbf{n}}(x,y,z)$ around the sphere with an accompanying hedgehog, results in a non-Newtonian response of LC, namely, shear thinning. Shear thinning has been observed in other nematics, such as a flow-aligning pentylcyanobiphenyl (5CB), whenever the shear direction is orthogonal to the director orientation imposed either by electric field [35-38], or by surface alignment [38,39]. When flow realigns the director towards the shear plane, the measured viscosity decreases, which is reasonable if the friction correlates with the cross-section of the rod-like molecules seen by the flow [30]. In a rheometer, shear thinning is expected if the shear aligns originally disorganized structure. Shear thinning is expected also for a moving colloid, since the director field $\hat{\mathbf{n}}(x,y,z)$ around it is not parallel to the far-field orientation $\hat{\mathbf{n}}_0 = (1,0,0)$ because of the perpendicular anchoring at the colloid's surface. This field exhibits components $\hat{n}_y \neq 0$, $\hat{n}_z \neq 0$ along the $y$- and $z$-axes, respectively, and a topological hedgehog defect, Fig.1(a-c) [22]. The effective viscosity should decrease when the director $\hat{\mathbf{n}}(x,y,z)$ realigns towards the direction of colloid's motion.

The viscosity $\eta$ of MLC7026 measured in a cone-plate rheometer shows a clear shear-thinning behavior for shear rates $\dot{\gamma} = (0.1 - 100)$ s$^{-1}$, Fig. 1(e). In this range, $\eta$ decreases from $\approx$ 46 mPa·s to $\approx$ 22.5 mPa·s, which is well fitted with $\eta = a\dot{\gamma}^{-0.7} + b$, where $a = 4.7$ mPa·s$^{0.3}$, and $b = 21.8$ mPa·s, Fig. 1(e). Of course, the actual value of $\eta$ measured in a rheometer is different from $\eta_{eff}$ in Eq. (2) characterizing the sphere-hedgehog director. We approximate the dependency $\eta_{\text{eff}}(v_x)$ using the measured $\eta(\dot{\gamma})$ data in Fig. 1(e), assuming that the shear rate $\dot{\gamma}$ produced by electrophoretic motion is $\dot{\gamma} = v_x/\left(\frac{h}{2}\right)$, where $h = 100$ μm is the cell thickness; the shear is created by the sphere moving in the midplane of the cell with respect to the immobilized plates. The field dependence of the product $v_x \eta_{eff}(v_x)$ is close to quadratic, $v_x \eta_{eff} = \mu^{(2)} E_z^2$, Fig. 1(f), as expected by Eq. (2). Some discrepancies between the experiment and the theoretical model are expected in light of rough estimates of the effective viscosities and shear rates. The shear thinning mechanism of the strong nonlinearity is also supported by the observation that at very high and very small fields, the speed follows the $v_x \propto E_z^2$ dependence, as expected in a conventional LCEP in the absence of the shear thinning effect and explained below.

At high fields, $E_z \geq 0.4$ V/µm, the velocity $v_x$ approaches and exceeds 100 µm/s, which means the shear rate $\dot{\gamma}$ is strong, on the order of 1 s$^{-1}$ and higher. The viscosity does not decrease much at these shear rates, Fig.1(e), which explains why $v_x \propto E_z^2$.

The quadratic dependence holds also at fields below 0.1 V/µm, at which $v_x \leq 0.1$ µm/s and $\dot{\gamma} \leq 2 \times 10^{-3}$ s$^{-1}$. The Ericksen number, which is the ratio of the viscous to elastic torques, $\mathrm{Er} = \frac{\eta \dot{\gamma}(h/2)^2}{K}$, is smaller than 1 at these rates; here $K$ is the average Frank elastic constant. With $\dot{\gamma} = 10^{-3}$ s$^{-1}$, $\eta = 46$ mPa·s (the maximum viscosity in Fig. 1(e)), $K = 10^{-11}$ $N$, one finds Er~0.01, which indicates that the elastic torques prevail and there is no significant shear thinning. Direct measurements of the effective viscosity of cyanobiphenyl nematics [35] confirm that $\eta$ does not depend on the shear rate when $\dot{\gamma} < 0.1$ s$^{-1}$. The absence of shear thinning suggests the conventional $v_x \propto E_z^2$ behavior at $E_z \leq 0.1$ $V/\mu m$, as observed in Fig. 1(d).

It is known that the LCEP speed decreases at high frequencies of the applied field [8]. A weaker speed implies a weaker shear rate. Experiments performed for MLC7026 and PS spheres of diameter 5 µm at frequencies 60 Hz and 300 Hz confirm that both the speed and the nonlinearity diminish, Supplementary Fig. S3; the exponent in the field dependency $v_x \propto |E_3|^\alpha$ decreases to the range $\alpha \approx 3 - 4$.

The strongly nonlinear regime $v_x \propto E_z^6$ is confined to a relatively narrow field range, $(0.20 - 0.35)$ V/µm. The corresponding velocities are on the order of $(2 - 40)$ µm/s, which implies shear rates $0.04 - 0.8$ s$^{-1}$. This range corresponds to the strongest shear-thinning: $\eta(\dot{\gamma})$ decreases dramatically in the range $\dot{\gamma} = (0.1 - 0.8)$ s$^{-1}$, Fig.1(e).

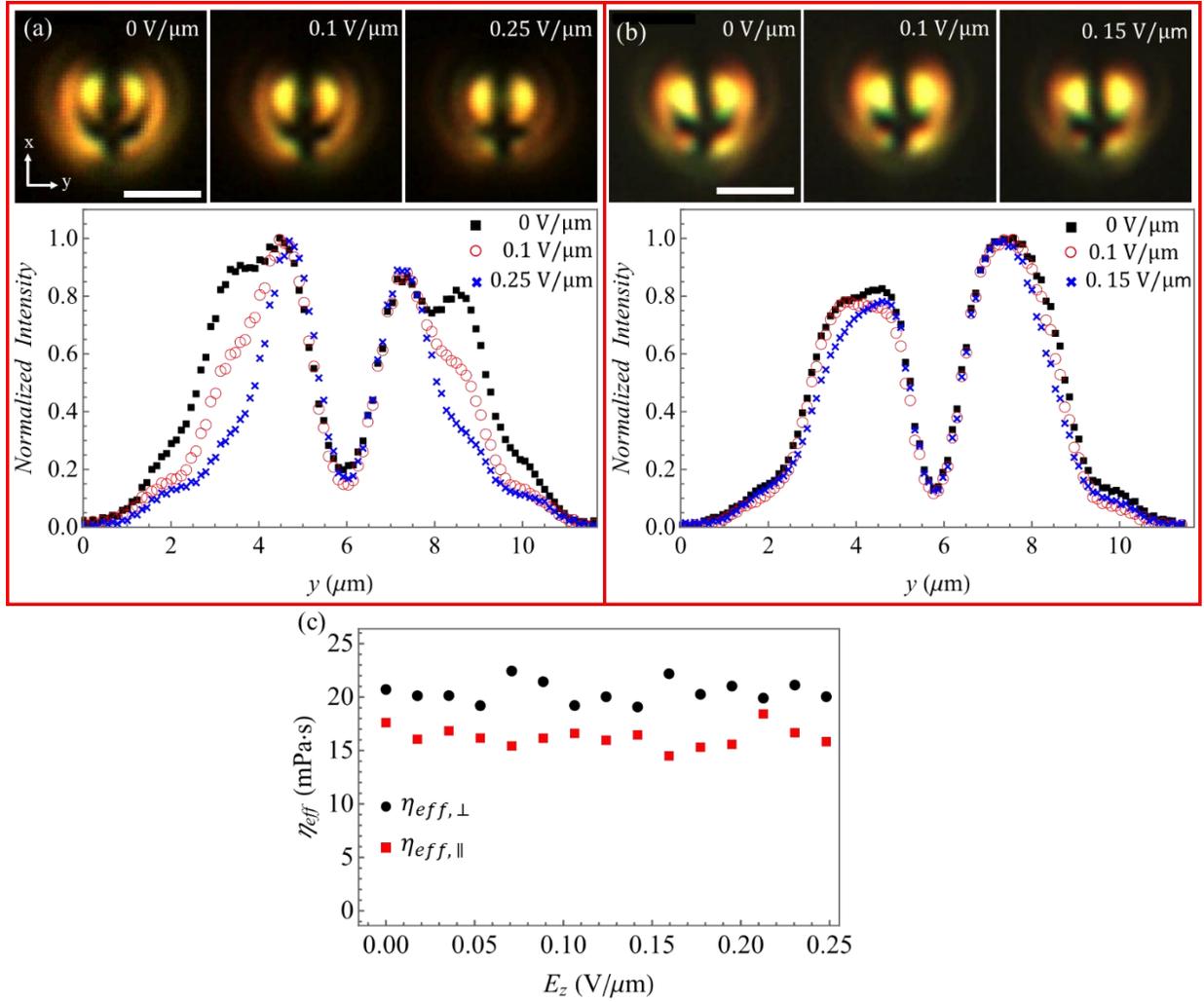

FIG. 2. Field induced director modifications. (a) Polarized optical microscopy textures of a PS sphere in MLC7026 ($\Delta\varepsilon = -3.7$) at different electric fields, and normalized transmitted intensity profile averaged over the *x*-axis; $f = 20$ Hz. (b) Polarized optical microscopy textures of a PS sphere in MLC7026+E7 ($|\Delta\varepsilon| < 10^{-3}$) at different fields, and normalized transmitted intensity profile; $f = 20$ Hz. The polarizer transmission axis is along the *x*-axis, and the analyzer is along the *y*-axis. (c) "Passive" viscosity of MLC7026 determined from the Brownian motion of PS spheres at different electric fields (frequency $f = 1$ kHz). The scale bars are 5 μm in (a) and (b); the camera exposure time is 0.5 s.

*Conclusion.* We demonstrated strongly nonlinear electrophoresis of colloidal spheres in a nematic electrolyte, in which the electrophoretic speed as the function of the driving ac electric field acquires fourth- and sixth-order terms. The effect is observed when the motion is sufficiently fast to create shear flow over an extended volume around the colloid. The reduced viscosity of LC due to shear leads to high nonlinearity. At low fields and low speeds, the shear thinning and nonlinearity are weak. On the other hand, very high fields and speeds reduce the nonlinearity since the viscosity dependence on the high shear rate becomes weaker. It would be of interest to explore cells in which the spheres are closer to one or both substrates and thus the effective shear rates are stronger. Also of interest is to use larger spheres, as the LCEP speed grows with the particles' size [8,21]. The explored nematic electrolytes are flow-aligning, as most of the thermotropic nematics formed by rod-like molecules [30]. An open question is whether a nonlinear electrophoresis could be observed in tumbling nematics, such as polymer [40] and chromonic lyotropic [41] solutions, or thermotropic nematics near the transition to the smectic phase [42].

To the best of our knowledge, this is the highest nonlinearity of an electrophoretic effect ever reported. The strong dependence on the applied electric field enhances the efficiency of liquid crystalline electrolytes for electrophoretic transport. Since the speed depends on the even-order powers of the electric field, the LCEP effect can be driven by an ac electric field, thus mitigating the detrimental effect of dc driving known in the classic electrophoresis observed in isotropic electrolytes.

**Acknowledgment**

The work was supported by NSF grants DMR-1905053 and DMR-2219151.

**Author contributions**: M.R. performed the experiments with the help of T.T. B.L. helped with the early-stage experiments. H.B. performed the shear experiment. M.R., D.G., and O.D.L. analyzed the data; M.R. and O.D.L. wrote the manuscript with input from all co-authors. O.D.L. conceived and supervised the project. All authors contributed to scientific discussions.

**Competing interests**: The authors declare no competing interests.

**Data Availability**: The data that support the plots within this paper and other findings of this study are available from the corresponding author.

**Supplementary Information**

**High order nonlinear electrophoresis in nematic liquid crystal**

Mojtaba Rajabi [1,2], Taras Turiv [1], Bing-Xiang Li [1,3], Hend Baza [1,2], Dmitry Golovaty [4], Oleg D. Lavrentovich [1,2,5,*]

[1]Advanced Materials and Liquid Crystal Institute, Kent State University, Kent, OH 44242, USA
[2]Department of Physics, Kent State University, Kent, OH 44242, USA
[3]College of Electronic and Optical Engineering & College of Flexible Electronics (Future Technology), Nanjing University of Posts and Telecommunications, Nanjing 210023, China
[4]Department of Mathematics, The University of Akron, Akron, OH 44325-4002, USA.
[5]Materials Science Graduate Program, Kent State University, Kent, OH 44242, USA
[*]Corresponding author. Email: olavrent@kent.edu

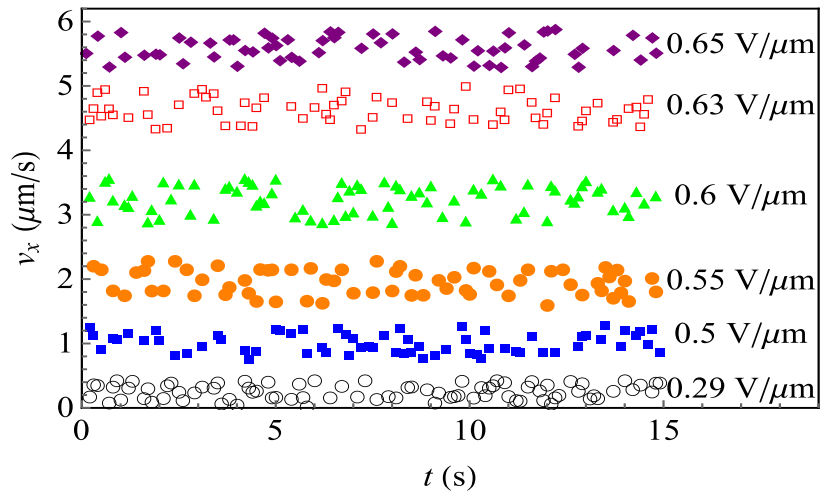

**Supplementary Fig. S1**. Electrophoretic speed of 5 µm diameter PS spheres as the function of time at different amplitudes of a sinusoidal ac electric field, $f = 20\ Hz$.

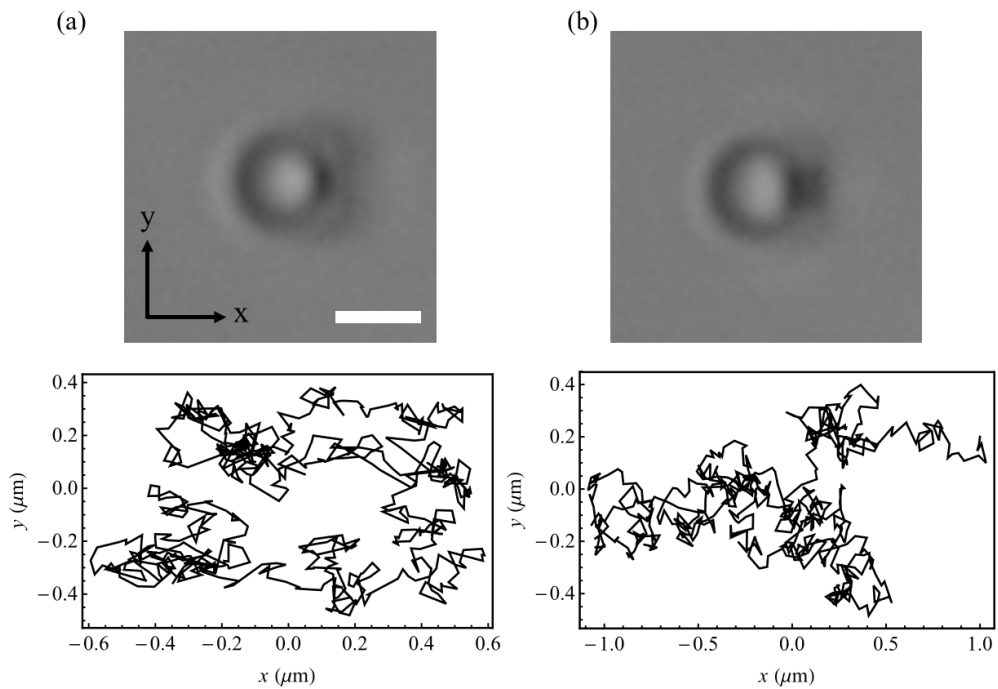

**Supplementary Fig. S2**. The optical texture and trajectory of a PS sphere at a) 0 V/μm, and b) 0.7 V/μm, $f = 1$ kHz. The scale bar is 5 μm.

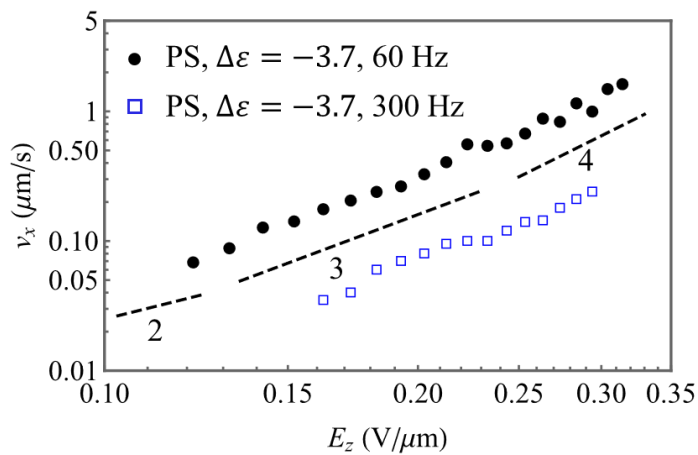

**Supplementary Fig. S3**. Electrophoretic speed of PS spheres in MLC7026 ($\Delta\varepsilon = -3.7$) as a function of applied ac electric field at $f = 60$ Hz and $f = 300$ Hz. The cell thickness $h = 200$ μm.